\documentclass[twocolumn,showpacs,preprintnumbers,amsmath,amssymb]{revtex4}
\usepackage{graphicx}% Include figure files
\usepackage{dcolumn}% Align table columns on decimal point
\usepackage{bm}% bold math

\usepackage{tikz}   % for LaTeX-drawn pictures. 
\usetikzlibrary{arrows}  % for fancier arrows in tikz

\pgfrealjobname{rss} % this is for extracting pgf/tikz figures:
\def\bk{\bm{k}}
\def\bq{\bm{q}}
\def\bx{\bm{x}}

\def\Qzz{Q_{zz}}
\def\Qzw{Q_{zw}}

\def\Qww{Q_{ww}}
\def\Qwz{Q_{wz}}

\begin{document}

\preprint{Revised preprint, November 2011}

\title{Three-wave interactions and spatio-temporal chaos}% Force line breaks with \\

\author{A. M. Rucklidge}
% \email{A.M.Rucklidge@leeds.ac.uk}
\affiliation{%
Department of Applied Mathematics, University of Leeds, Leeds LS2 9JT, UK}%

\author{M. Silber}
% \email{m-silber@northwestern.edu}
\affiliation{%
Department of Engineering Sciences and Applied Mathematics, and               
% Northwestern Institute on Complex Systems,
 NICO,
Northwestern University, Evanston, IL 60208, USA}%

\author{A. C. Skeldon}
% \email{A.Skeldon@surrey.ac.uk}
\affiliation{%
Department of Mathematics,
University of Surrey,                                           
Guildford GU2 7XH, UK}                                     

\date{2 November 2011}% It is always \today, today,

\begin{abstract}
Three-wave interactions form the basis of our understanding of many pattern
forming systems because they encapsulate the most basic nonlinear interactions.
In problems with two comparable length scales, it is possible for two waves of
the shorter wavelength to interact with one wave of the longer, as well as for
two waves of the longer wavelength to interact with one wave of the shorter.
Consideration of both types of three-wave interactions can generically explain
the presence of complex patterns and spatio-temporal chaos. Two length scales
arise naturally in the Faraday wave experiment with multi-frequency forcing,
and our results enable some previously unexplained experimental observations of
spatio-temporal chaos to be interpreted in a new light. Our predictions are
illustrated with numerical simulations of a model partial differential
equation.

\end{abstract}


% 47.54.-r Pattern formation in fluid dynamics
% 47.52.+j Chaos in fluid dynamics
% 47.10.Fg Dynamical Systems methods in Fluid Dynamics
% 05.45.-a Nonlinear Dynamics/bifurcation
% 89.75.Kd Pattern formation in complex systems
% 47.20.Ma Interfacial flow instabilities
% 47.35.Pq Capillary waves, fluid flow

\maketitle

Patterns arise in many non-equilibrium physical, chemical and
biological systems, often when a uniform state is subjected to external driving
and becomes unstable to modes with a finite wavelength. In some systems,
modes with a second wavelength can play an important role in pattern formation
if these modes are either unstable or only weakly damped. The
interaction between two waves of one wavelength with a third wave of the other
wavelength is known 
both experimentally and theoretically
to play a key role in producing a rich variety of
interesting phenomena such as quasipatterns, superlattice patterns and
spatio-temporal chaos
(STC)~\cite{Arbell2002Epstein2004Epstein2006Kudrolli1998,
  Berenstein2004,
  Ding2006,
  Edwards1994,Porter2004,Zhang1996,
  Rogers2005Rucklidge2009Topaz2004,
  Ackemann2001Vorontsov1998}.

In this paper, we focus on three-wave interactions (3WIs) involving two comparable
wavelengths and develop a criterion for when such interactions are likely to
lead to~\hbox{STC}, as opposed to steady patterns and quasipatterns.  
The mechanism we describe is generic, and will apply
to any system in which such 3WIs can occur, such as the
Faraday wave 
experiment~\cite{Arbell2002Epstein2004Epstein2006Kudrolli1998,Edwards1994,Ding2006}, 
coupled Turing systems~\cite{Berenstein2004} 
and some optical systems~\cite{Ackemann2001Vorontsov1998}. In
order to illustrate our criterion we focus on two examples.  The first
is the Faraday experiment, in which patterns of standing waves are excited on
the surface of a fluid by periodically forcing the fluids' container up and
down. Using a multi-component forcing $f(t)$ enables the excitation of waves
with comparable wavelengths.  Experimentally, the phases and amplitudes of the
different components of $f(t)$ have been shown to determine whether
simple pattterns, superlattice patterns, quasipatterns 
or~\hbox{STC} are 
seen~\cite{Arbell2002Epstein2004Epstein2006Kudrolli1998,Edwards1994,Ding2006}. 
A theoretical understanding of the stabilization of some superlattice patterns
has been developed using a single 3WI~\cite{Edwards1994,Porter2004,Zhang1996}, but up until this
point there has been no explanation for the presence of~\hbox{STC}.  We show
how our extension of the notion of 3WIs can explain the
origin of STC, quasipatterns and other features of the experimental results
of~\cite{Ding2006}. Our second example is a new model PDE that has been
designed to enable an exploration of 3WIs with two
comparable wavelengths.
\begin{figure}
\makebox[\hsize]{%
% scale=3: basic units are cm, so -1..1 is 2cm wide, scale=3 brings this to 6cm
% >=stealth produces nicer shaped arrows
\mbox{\beginpgfgraphicnamed{rss_fig_threewaves_kkq}%
\begin{tikzpicture}[scale=1.00,>=stealth]
%   \useasboundingbox (-1.15,-1.15) rectangle (1.5,1.6);
   \draw (-1,1.1) node[above] {(a)};
% drawing area
% \draw[thin] (-1,-1) -- (-1,1) -- (1,1) -- (1,-1) -- (-1,-1);
%axes
   \draw[->] (-1.2,0) -- (1.2,0) node[right=-2pt] {$k_x$}; 
   \draw[->] (0,-1.2) -- (0,1.2) node[above=-2pt] {$k_y$};       
% circle k=1
   \draw[thick] (0,0) circle (1.0);
% circle k=q=0.6600
   \draw[thick] (0,0) circle (0.6600);
% Coordinates of z vectors: (level 0)
 \draw[thick,->] (0,0) -- (  0.3300,  0.9440) node[above=2pt,right=0pt] {$\bk_1$};
 \draw[thick,->] (0,0) -- ( -0.3300,  0.9440);
 \draw[thick,->] (0,0) -- ( -0.3300, -0.9440);
 \draw[thick,->] (0,0) -- (  0.3300, -0.9440) node[above=-6pt,right=0pt] {\smash{$\bk_2$}};
 \draw[thick,->] ( 0.3300,  0.9440) -- (  0.6600,  0.0000);
% Coordinates of w vectors: (level 0)
 \draw[thick,->] (0,0) -- (  0.6600,  0.0000) node[above=4pt,right=-3.0pt] {$\bq_1$};
 \draw[thick,->] (0,0) -- ( -0.6600,  0.0000);
% arc for theta_z
   \draw (0.3,0) arc (0:70.73:0.3);
   \draw (0.3,0) arc (0:-70.73:0.3);
   \draw (0.3,0) arc (0:35:0.3) -- (0.95,0.63); 
   \draw (1.0,0.63) node[above=0pt, right=-3pt] {$\theta_z$};
\end{tikzpicture}\endpgfgraphicnamed}%
\hfil%
\mbox{\beginpgfgraphicnamed{rss_fig_threewaves_kqq}%
\begin{tikzpicture}[scale=1.00,>=stealth]
%   \useasboundingbox (-1.15,-1.15) rectangle (1.5,1.6);
   \draw (-1,1.1) node[above] {(b)};
% drawing area
% \draw[thin] (-1,-1) -- (-1,1) -- (1,1) -- (1,-1) -- (-1,-1);
%axes
   \draw[->] (-1.2,0) -- (1.2,0) node[right=-2pt] {$k_x$}; 
   \draw[->] (0,-1.2) -- (0,1.2) node[above=-2pt] {$k_y$};       
% circle k=1
   \draw[thick] (0,0) circle (1.0);
% circle k=q=0.6600
   \draw[thick] (0,0) circle (0.6600);
% Coordinates of z vectors: (level 0)
 \draw[thick,->] (0,0) -- (  0.3300,  0.9440) node[above=2pt,right=0pt] {$\bk_1$};
 \draw[thick,->] (0,0) -- ( -0.3300, -0.9440);
% this is to help keep everything aligned: an invisible k_2
 \draw                    (  0.3300, -0.9440) node[above=-6pt,right=0pt] {\phantom{\smash{$\bk_2$}}};
% Coordinates of w vectors: (level 0)
 \draw[thick,->] (0,0) -- (  0.5717,  0.3298) node[above=1pt,right=-3pt] {$\bq_2$};
 \draw[thick,->] (0,0) -- ( -0.2417,  0.6142) node[above=-2pt] {$\bq_3$};
 \draw[thick,->] (0,0) -- ( -0.5717, -0.3298);
 \draw[thick,->] (0,0) -- (  0.2417, -0.6142);
 \draw[thick,->] (  0.5717,  0.3298) -- (  0.3300,  0.9440);
 \draw (0.2599,0.1499) arc (29.98:111.48:0.3);
 \draw (0.2599,0.1499) arc (29.98:50:0.3) -- (0.90,0.90);
 \draw (0.95,0.90) node[above=0pt, right=-3pt] {$\theta_w$};
% \draw (0.75,1.15) node {$\theta_w=2\arccos\left(\frac{1}{2q}\right)$};
\end{tikzpicture}\endpgfgraphicnamed}%
\hfil%
\mbox{\beginpgfgraphicnamed{rss_fig_threewaves_lots}%
\begin{tikzpicture}[scale=1.00,>=stealth]
%   \useasboundingbox (-1.15,-1.15) rectangle (1.5,1.6);
   \draw (-1,1.1) node[above] {(c)};
% drawing area
% \draw[thin] (-1,-1) -- (-1,1) -- (1,1) -- (1,-1) -- (-1,-1);
%axes
   \draw[->] (-1.2,0) -- (1.2,0) node[right=-2pt] {$k_x$}; 
   \draw[->] (0,-1.2) -- (0,1.2) node[above=-2pt] {$k_y$};       
% circle k=1
   \draw[thick] (0,0) circle (1.0);
% circle k=q=0.66
   \draw[thick] (0,0) circle (0.6600);
% Coordinates of z vectors: (level 0)
 \draw[thick,->] (0,0) -- (  0.3300,  0.9440) node[above=2pt,right=0pt] {$\bk_1$};
 \draw[thick,->] (0,0) -- ( -0.3300,  0.9440);
 \draw[thick,->] (0,0) -- ( -0.3300, -0.9440);
 \draw[thick,->] (0,0) -- (  0.3300, -0.9440) node[above=-6pt,right=0pt] {\smash{$\bk_2$}};
% Coordinates of w vectors: (level 0)
 \draw[thick,->] (0,0) -- (  0.6600,  0.0000) node[above=4pt,right=-3.0pt] {$\bq_1$};
 \draw[thick,->] (0,0) -- ( -0.6600,  0.0000);
% Coordinates of z vectors: (level       1)
 \draw[->] (0,0) -- (  0.9993,  0.0386);
 \draw[->] (0,0) -- (  0.7576,  0.6527);
 \draw[->] (0,0) -- (  0.3300,  0.9440);
 \draw[->] (0,0) -- (  0.1859,  0.9826);
 \draw[->] (0,0) -- ( -0.1859,  0.9826);
 \draw[->] (0,0) -- ( -0.3300,  0.9440);
 \draw[->] (0,0) -- ( -0.7576,  0.6527);
 \draw[->] (0,0) -- ( -0.9993,  0.0386);
 \draw[->] (0,0) -- ( -0.9993, -0.0386);
 \draw[->] (0,0) -- ( -0.7576, -0.6527);
 \draw[->] (0,0) -- ( -0.3300, -0.9440);
 \draw[->] (0,0) -- ( -0.1859, -0.9826);
 \draw[->] (0,0) -- (  0.1859, -0.9826);
 \draw[->] (0,0) -- (  0.3300, -0.9440);
 \draw[->] (0,0) -- (  0.7576, -0.6527);
 \draw[->] (0,0) -- (  0.9993, -0.0386);
% Coordinates of w vectors: (level       1)
 \draw[->] (0,0) -- (  0.6600,  0.0000);
 \draw[->] (0,0) -- (  0.5717,  0.3298) node[above=1pt,right=-3pt] {$\bq_2$};
 \draw[->] (0,0) -- (  0.2417,  0.6142);
 \draw[->] (0,0) -- ( -0.2417,  0.6142) node[above=-2pt] {$\bq_3$};
 \draw[->] (0,0) -- ( -0.5717,  0.3298);
 \draw[->] (0,0) -- ( -0.6600,  0.0000);
 \draw[->] (0,0) -- ( -0.5717, -0.3298);
 \draw[->] (0,0) -- ( -0.2417, -0.6142);
 \draw[->] (0,0) -- (  0.2417, -0.6142);
 \draw[->] (0,0) -- (  0.5717, -0.3298);
% Coordinates of z vectors: (level       2)
 \draw[->] (0,0) -- (  0.9993,  0.0386);
 \draw[->] (0,0) -- (  0.9824,  0.1868);
 \draw[->] (0,0) -- (  0.8848,  0.4659);
 \draw[->] (0,0) -- (  0.8463,  0.5328);
 \draw[->] (0,0) -- (  0.7576,  0.6527);
 \draw[->] (0,0) -- (  0.6520,  0.7582);
 \draw[->] (0,0) -- (  0.4018,  0.9157);
 \draw[->] (0,0) -- (  0.3300,  0.9440);
 \draw[->] (0,0) -- (  0.1859,  0.9826);
 \draw[->] (0,0) -- ( -0.1859,  0.9826);
 \draw[->] (0,0) -- ( -0.3300,  0.9440);
 \draw[->] (0,0) -- ( -0.4018,  0.9157);
 \draw[->] (0,0) -- ( -0.6520,  0.7582);
 \draw[->] (0,0) -- ( -0.7576,  0.6527);
 \draw[->] (0,0) -- ( -0.8463,  0.5328);
 \draw[->] (0,0) -- ( -0.8848,  0.4659);
 \draw[->] (0,0) -- ( -0.9824,  0.1868);
 \draw[->] (0,0) -- ( -0.9993,  0.0386);
 \draw[->] (0,0) -- ( -0.9993, -0.0386);
 \draw[->] (0,0) -- ( -0.9824, -0.1868);
 \draw[->] (0,0) -- ( -0.8848, -0.4659);
 \draw[->] (0,0) -- ( -0.8463, -0.5328);
 \draw[->] (0,0) -- ( -0.7576, -0.6527);
 \draw[->] (0,0) -- ( -0.6520, -0.7582);
 \draw[->] (0,0) -- ( -0.4018, -0.9157);
 \draw[->] (0,0) -- ( -0.3300, -0.9440);
 \draw[->] (0,0) -- ( -0.1859, -0.9826);
 \draw[->] (0,0) -- (  0.1859, -0.9826);
 \draw[->] (0,0) -- (  0.3300, -0.9440);
 \draw[->] (0,0) -- (  0.4018, -0.9157);
 \draw[->] (0,0) -- (  0.6520, -0.7582);
 \draw[->] (0,0) -- (  0.7576, -0.6527);
 \draw[->] (0,0) -- (  0.8463, -0.5328);
 \draw[->] (0,0) -- (  0.8848, -0.4659);
 \draw[->] (0,0) -- (  0.9824, -0.1868);
 \draw[->] (0,0) -- (  0.9993, -0.0386);
% Coordinates of w vectors: (level       2)
 \draw[->] (0,0) -- (  0.6600,  0.0000);
 \draw[->] (0,0) -- (  0.5717,  0.3298);
 \draw[->] (0,0) -- (  0.5163,  0.4112);
 \draw[->] (0,0) -- (  0.4830,  0.4498);
 \draw[->] (0,0) -- (  0.3304,  0.5714);
 \draw[->] (0,0) -- (  0.2417,  0.6142);
 \draw[->] (0,0) -- (  0.0976,  0.6527);
 \draw[->] (0,0) -- ( -0.0976,  0.6527);
 \draw[->] (0,0) -- ( -0.2417,  0.6142);
 \draw[->] (0,0) -- ( -0.3304,  0.5714);
 \draw[->] (0,0) -- ( -0.4830,  0.4498);
 \draw[->] (0,0) -- ( -0.5163,  0.4112);
 \draw[->] (0,0) -- ( -0.5717,  0.3298);
 \draw[->] (0,0) -- ( -0.6600,  0.0000);
 \draw[->] (0,0) -- ( -0.5717, -0.3298);
 \draw[->] (0,0) -- ( -0.5163, -0.4112);
 \draw[->] (0,0) -- ( -0.4830, -0.4498);
 \draw[->] (0,0) -- ( -0.3304, -0.5714);
 \draw[->] (0,0) -- ( -0.2417, -0.6142);
 \draw[->] (0,0) -- ( -0.0976, -0.6527);
 \draw[->] (0,0) -- (  0.0976, -0.6527);
 \draw[->] (0,0) -- (  0.2417, -0.6142);
 \draw[->] (0,0) -- (  0.3304, -0.5714);
 \draw[->] (0,0) -- (  0.4830, -0.4498);
 \draw[->] (0,0) -- (  0.5163, -0.4112);
 \draw[->] (0,0) -- (  0.5717, -0.3298);
\end{tikzpicture}\endpgfgraphicnamed}}

\vspace{-1.0ex}

\caption{\label{fig:threewaves}Three-wave interactions between waves on two 
critical circles, with outer radius~$1$ and inner radius 
$q>\frac{1}{2}$. 
 (a)~A vector $\bq_1$ on the inner circle can be written as the sum of two
vectors $\bk_1$ and $\bk_2$ on the outer circle; this defines the angle
$\theta_z=2\arccos(q/2)$.
 (b)~The vector $\bk_1$ is the sum of two inner vectors $\bq_2$ and $\bq_3$; 
this defines the angle $\theta_w=2\arccos(1/2q)$.
 (c)~Similarly, $\bk_2$ is the sum of two inner vectors, and
each of these is in turn the sum of two outer vectors, and so on.}
%, and so on.
%In this example, $q=0.66$, so $\theta_z=141.4^\circ$ and
%$\theta_w=81.5^\circ$.}
 \end{figure}
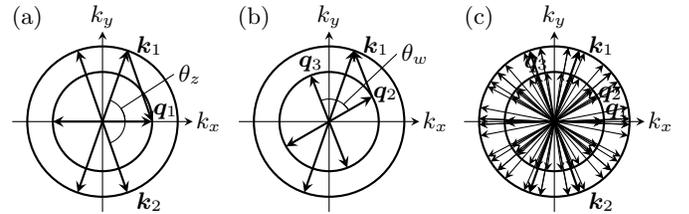
%%%%%%%%%%%%%%%%%%%%%%%%%%%%%%%%%%%%%%%%%%%%%%%%%%%%%%%%%%%%%%%%%%%%%%%%%%%%%%%%%

We suppose that in the system two wavenumbers $k=1$ and $k=q$ are excited with
$0<q<1$. 3WIs then can take two forms: first, two waves on
the outer circle (wavenumber~$1$) interact nonlinearly with a wave on the
inner circle (wavenumber~$q$), for example $\bk_1$ and $\bk_2$ with
$\bq_1=\bk_1+\bk_2$ (figure~\ref{fig:threewaves}a); and second, if
$\frac{1}{2}<q<1$, two waves on the inner circle interact with one on the outer
circle, for example vectors $\bq_2$ and $\bq_3$ with $\bk_1=\bq_2+\bq_3$
(figure~\ref{fig:threewaves}b). The consequence of the presence of both forms
of interaction when $\frac{1}{2}<q<1$ is that, for almost all choices of~$q$,
nonlinear interactions necessarily generate
infinitely many modes on both critical circles 
(figure~\ref{fig:threewaves}c). This occurs because each vector on the outer
circle is the sum of two inner vectors, and in turn each of these inner vectors
is the sum of two (different) outer vectors, and so on.
 \looseness=-1

% {\it ad infinitum}.

So, for $\frac{1}{2}<q<1$, each mode on each circle will
be influenced by two types of 3WI: one type involving    
interaction with a pair of modes from the other circle, and the second
involving interaction with one mode from each circle.  (There is also the
usual interaction between three waves at $120^\circ$ on a single circle.)
 To understand the effect of this multitude of nonlinear interactions
(figure~\ref{fig:threewaves}c),
we write the 
pattern-forming field~$U(x,y,t)$ as
 $$
 U=\sum_{j=1, |\bk_j|=1}^{\infty}z_j(t)e^{i\bk_j\cdot\bx} + 
   \sum_{j=1, |\bq_j|=q}^{\infty}w_j(t)e^{i\bq_j\cdot\bx} + \text{h.o.t.},
 $$
where $t$ is the time-scale for the slow evolution of the amplitudes,
and the sums are over all modes on the two circles.
In order to quantify the four coefficients that govern the two 
types of 3WIs, we
write down archetypal equations (at lowest order) for the two
different types of 3WI, first involving a triad of
wavevectors $\bk_1$,
$\bk_2$ and $\bq_1=\bk_1+\bk_2$, with amplitudes $z_1$, $z_2$ and~$w_1$:
 \begin{eqnarray}
 \dot{z}_1 =& \mu z_1 + \Qzw \bar{z}_2 w_1 + \text{cubic terms}, \nonumber \\ 
 \dot{z}_2 =& \mu z_2 + \Qzw \bar{z}_1 w_1 + \text{cubic terms},  \label{eq:zzw} \\
 \dot{w}_1 =& \nu w_1 + \Qzz z_1 z_2 + \text{cubic terms}, \nonumber
 \end{eqnarray} 
where $\mu$ and $\nu$ are growth rates corresponding to wave\-numbers $1$ and~$q$, and
$\Qzw$ and $\Qzz$ are coefficients of the quadratic terms. Cubic terms play an
important role in the dynamics 
of~(\ref{eq:zzw})~\cite{Guckenheimer1992bPorter2004a}, and their
effect is included in the discussion below. We write similar equations for the second
type of 3WI, involving the triad $\bq_2$, $\bq_3$ and
$\bk_1=\bq_2+\bq_3$, with amplitudes $w_2$, $w_3$ and~$z_1$:
 \begin{eqnarray}
 \dot{w}_2 =& \nu w_2 + \Qwz \bar{w}_3 z_1 + \text{cubic terms},  \nonumber \\
 \dot{w}_3 =& \nu w_3 + \Qwz \bar{w}_2 z_1 + \text{cubic terms},  \label{eq:zww}  \\
 \dot{z}_1 =& \mu z_1 + \Qww  w_2 w_3 + \text{cubic terms}, \nonumber
 \end{eqnarray} 
where $\Qwz$ and $\Qww$ are two more quadratic coefficients.

These equations should be replicated for all possible
combinations of modes from the two circles, whenever 3WI is
possible. The
evolution of this infinite set of waves could be exceedingly complicated, but
we argue it is likely to be most strongly influenced by the outcome of the two
types of 3WIs governed by (\ref{eq:zzw}) and~(\ref{eq:zww}). To
this end, we summarize the results of~\cite{Guckenheimer1992bPorter2004a} and
others on equations~(\ref{eq:zzw}) with appropriate cubic terms. Pure mode
solutions are stable for
ranges of values of $\mu$ and~$\nu$. The behavior of mixed solutions, where
all three amplitudes are non-zero, is influenced heavily by the signs of the
quadratic coefficients. If these have the same sign
($\Qzz\Qzw>0$), there is a range of values of $(\mu,\nu)$ where steady
solutions of~(\ref{eq:zzw}) with all three modes non-zero are stable. In
contrast, if the quadratic coefficients have opposite sign ($\Qzz\Qzw<0$),
there is a range of values of $(\mu,\nu)$ where there is time-dependent
competition between the three amplitudes~\cite{Guckenheimer1992bPorter2004a}.
 \looseness=-1

In a similar manner,  in the multitude of 3WIs that occur
when $\frac{1}{2}<q<1$, the behavior of each type of 3WI
will depend on the products of the quadratic terms, $\Qzz\Qzw$ and $\Qww\Qwz$
in (\ref{eq:zzw}) and~(\ref{eq:zww}), and will determine whether each type of
interaction results in competition between modes or not.  Consequently the
signs of $\Qzz\Qzw$ and $\Qww\Qwz$ will strongly influence the dynamics
of the overall system.  Specifically, if $\Qzz\Qzw$ and $\Qww\Qwz$
are both negative, then we expect to find time-dependent competition leading to
intermittent appearance of patterns with $\theta_z$ or $\theta_w$, including
the possibility of~\hbox{STC}, with two full circles of modes present.
Conversely, if $\Qzz\Qzw$ and $\Qww\Qwz$ are both positive, only steady
patterns are to be expected: these may be
simple patterns such as stripes or hexagons, or they may 
have complex spatial
structure, with a superposition of waves having a variety of
orientations featuring the special angles $\theta_w$ and $\theta_z$, possibly 
with patches of one pattern in a background of the other. In the
intermediate case where, $\Qzz\Qzw>0$ and $\Qww\Qwz<0$ (or the other way
around), then resonant triad interactions involving $\theta_z$ are reinforced,
while triads involving $\theta_w$ can be competitive (or vice-versa). In this
case we expect to find steady superlattice patterns involving~$\theta_z$, or
time-dependent
competition between such superlattice patterns with different orientations.
 % \looseness=-1

We emphasise that 
these expectations for the overall behavior of the pattern-forming problem are
based on the idea that the dynamics of any particular vector drawn from the
infinite number of vectors on the two critical circles will be dominated by its
linear growth rate and the 3WIs associated with the two
types of triad to which that vector belongs.

There are two special values of the radius ratio~$q$ that do not immediately
generate an infinite number of waves: when $q$ is
$\frac{1}{2}(\sqrt{5}-1)=0.6180$ or $\frac{1}{2}(\sqrt{6}-\sqrt{2})=0.5176$,
corresponding to $\theta_z=2\theta_w=144^\circ$ and
$\theta_z=5\theta_w=150^\circ$, only a finite number of wavevectors on the
critical circles is generated. These two special cases correspond to 10-fold
and 12-fold quasipatterns respectively (although an infinite set of vectors is
generated in the 10-fold case once hexagonal interactions are included).
 The values of~$q$ that would lead to 8-fold patterns generate an
infinite number of vectors. Hexagons and squares are also not exceptions: for
example, if $q=1/\sqrt{2}$, we have $\theta_w=90^\circ$ but
$\theta_z=138.6^\circ$, which does not fit in a square lattice.

%%%%%%%%%%%%%%%%%%%%%%%%%%%%%%%%%%%%%%%%%%%%%%%%%%%%%%%%%%%%%%%%%%%%%%%%%%%%%%%%%
% Figure showing Ding and Umbanhowar's results
% the box in (a) is 893x723   = 1.235 x 1
% the box in (b) is 1120x530  = 2.113 x 1
% the ratio of these is 1.71, hence the ratio of the numbers below
\begin{figure*}
\def\framebox#1{{#1}}%
\makebox[\hsize]{%
% These two sizes are calculated to give figures that are the same height
% numbers that work (approximately) are:
% 100%: 0.369 and 0.631
%  90%: 0.332 and 0.568
%  80%: 0.295 and 0.505  raise 1.55truein
%  75%: 0.277 and 0.473
%  70%: 0.258 and 0.442  
%  60%: 0.221 and 0.379
\hfil\framebox{\raisebox{1.45truein}{(a)}}%
\framebox{\includegraphics[width=0.276\hsize]{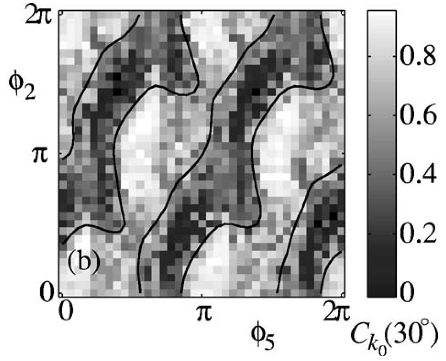}}%
\hfil\framebox{\raisebox{1.45truein}{(b)}}%                      
\framebox{\includegraphics[width=0.472\hsize]{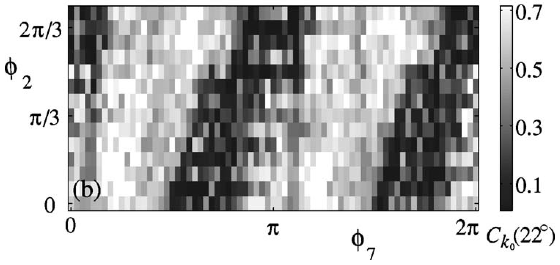}}}%

\makebox[\hsize]{%
\hfil\framebox{\raisebox{1.40truein}{(c)}}%
\framebox{\includegraphics[width=0.22\hsize]{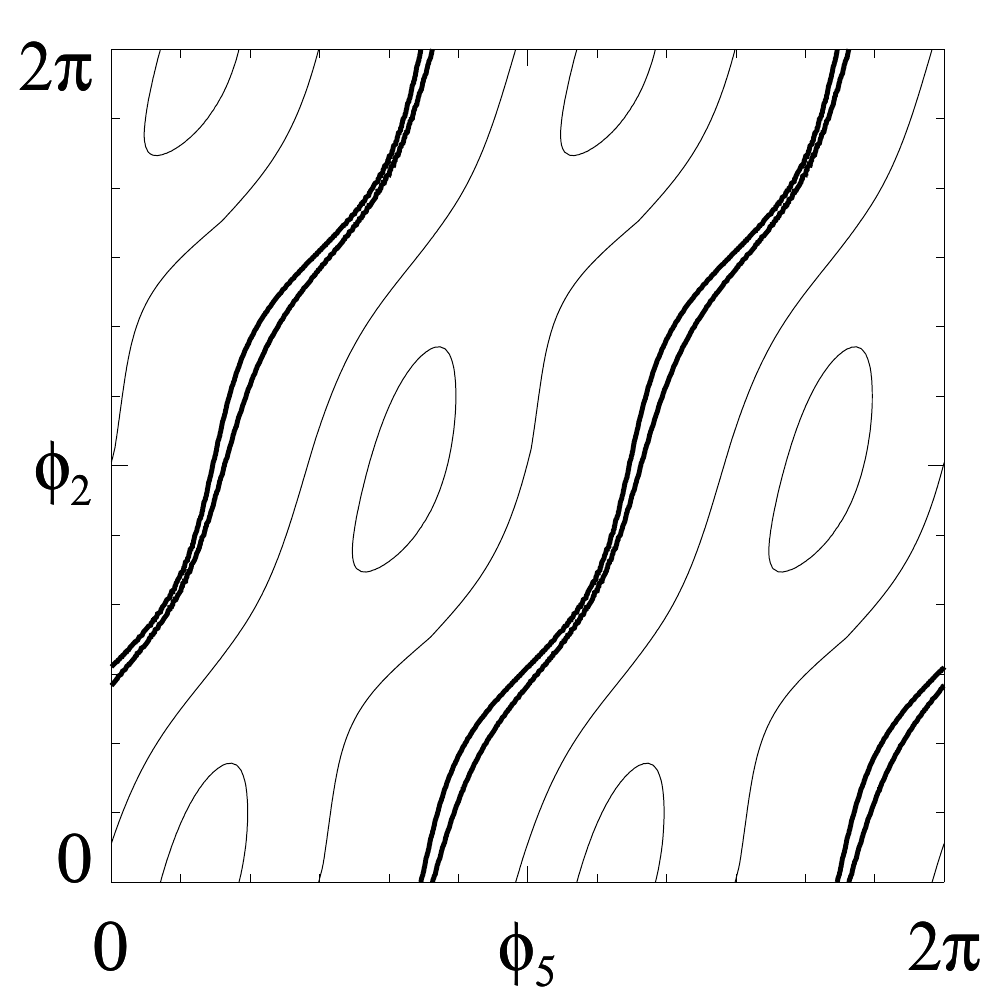}}
\hfil\framebox{\raisebox{1.40truein}{(d)}}%
\framebox{\includegraphics[width=0.22\hsize]{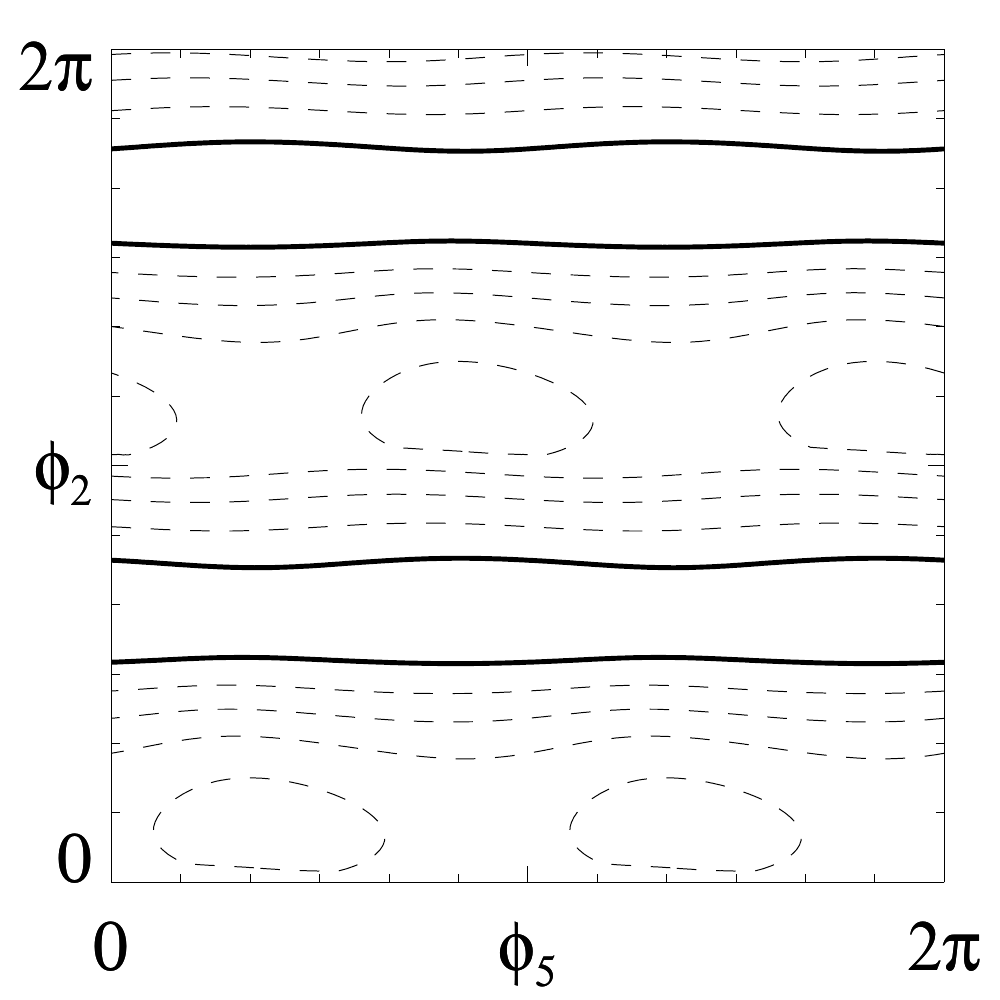}}
\hfil\framebox{\raisebox{1.40truein}{(e)}}%
\framebox{\includegraphics[width=0.22\hsize]{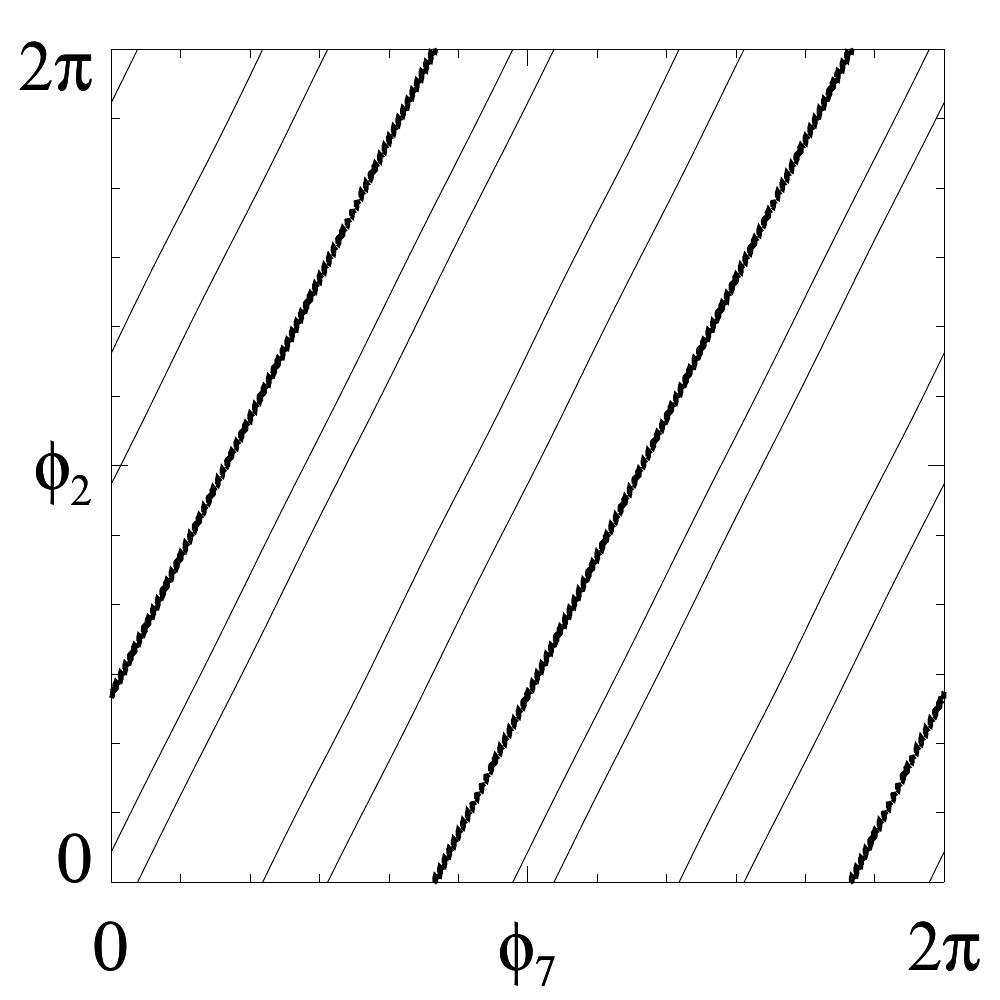}}}

\vspace{-2.0ex}

\caption{\label{fig:DUresults}(a,b)~Experimental results of Ding and 
Umbanhowar~\cite{Ding2006}, reproduced with permission. 
(a)~The $30^\circ$~angular autocorrelation function with $(4,5,2)$ 
forcing: white indicates
12-fold quasipatterns, as a function of~$(\phi_5,\phi_2)$. 
Black indicates disordered patterns.
(b)~The $22^\circ$~angular autocorrelation function with $(6,7,2)$ 
forcing: white indicates 
$22^\circ$~superlattice patterns, as a function of~$(\phi_7,\phi_2)$.
(c,d,e)~Products of quadratic coefficients.
(c,d)~Parameter values as in the experiment in~(a), with
(c)~$\Qzz\Qzw$ (contours in steps of $0.2$) and 
(d)~$\Qww\Qwz$ (contours in steps of $0.05$)
(e)~Parameter values as in~(b), showing
$\Qzz\Qzw$ (contours in steps of $0.1$). 
In all cases, the thick line is the zero contour, solid (dashed) 
lines are positive (negative)
contours. 
In~(c,e), $\Qzz\Qzw$ is negative only in very narrow diagonal bands; 
in~(d), $\Qww\Qwz$ is positive in two horizontal bands.}
\end{figure*}
%%%%%%%%%%%%%%%%%%%%%%%%%%%%%%%%%%%%%%%%%%%%%%%%%%%%%%%%%%%%%%%%%%%%%%%%%%%%%%%%%

Considering both types of 3WIs provides an explanation of
some of the observed Faraday wave phenomena in the experiments
of~\cite{Ding2006}. These use the three-frequency forcing
 $
 f(t) = a_m \cos \left ( m \omega t \right) +
        a_n \cos \left ( n \omega t + \phi_n \right) +
        a_p \cos \left ( p \omega t + \phi_p \right),
 $
where $(m,n,p)=(4,5,2)$ and $(6,7,2)$. The
3WIs are between harmonic modes, driven by the $(4,2)$ and
$(6,2)$ components of the forcing, with the primary instability being to the
larger frequency. The radius ratios are $q\approx0.52$ and $q\approx0.38$ in
the two cases: these are the radius ratios that arise in 12-fold quasipatterns
and $22^\circ$~superlattice patterns. The presence of these two patterns as a
function of the phases $(\phi_5,\phi_2)$ and $(\phi_7,\phi_2)$ is indicated by
light areas in figure~\ref{fig:DUresults}(a,b). The slopes of the bands and
their periodicity are related to time translation symmetries of components
of~$f(t)$~\cite{Porter2004}.

The two cases, $(4,5,2)$ and $(6,7,2)$ excitation, differ in that in the first
case, $q>\frac{1}{2}$, and `two-way' 3WIs between the different critical
modes is possible, whereas in the second case, $q<\frac{1}{2}$, and no such
two-way 3WIs can occur. We have calculated the values of the relevant
quadratic coefficients using weakly nonlinear analysis of the Navier--Stokes
equations for the experimental parameters by extending the method
in~\cite{Skeldon2007}. In figure~\ref{fig:DUresults}(c,d), we show $\Qzz\Qzw$
and $\Qww\Qwz$ for the $(4,5,2)$ case. Both products can take either positive
or negative values, although each is dominated by one sign.
In figure~\ref{fig:DUresults}(e), we show $\Qzz\Qzw$ for the $(6,7,2)$
case. Comparing with the experimental results in
figure~\ref{fig:DUresults}(a,b), we note that regions of 12-fold quasipatterns
and $22^\circ$~superlattice patterns correlate extremely well with regions of
positive~$\Qzz\Qzw$. Conversely, when $\Qzz\Qzw$ is small or negative, the
autocorrelation function in the experiments is low and STC is
seen~\cite{Ding2011}. The extra structure in figure~\ref{fig:DUresults}(a),
where the dark stripes broaden horizontally just above $\phi_2=\frac{\pi}{2}$ and 
$\frac{3\pi}{2}$, 
is aligned with changes in $\Qww\Qwz$ in
figure~\ref{fig:DUresults}(d), although our expectation would be for enhanced
STC where $\Qww\Qwz$ is negative.

There is clearly scope for further investigation of the role of the two-way 
3WIs in this and other experiments, motivated by the
encouraging alignment between the experimental autocorrelation functions and the
calculated quadratic coefficients.

We have further investigated the importance of the signs of the quadratic 
coefficients in the two-way 3WIs by devising a new model PDE for a field
$U(x,y,t)$:
 \begin{equation}
 \frac{\partial U}{\partial t} = {\cal{L}}(\mu,\nu) U + Q_1U^2 + Q_2U\nabla^2U + 
 Q_3\left|\nabla U\right|^2 - U^3.
 \label{eq:PDE}
 \end{equation}
The model (\ref{eq:PDE}) is an extension of those of~\cite{Swift1977Lifshitz1997},
modified to allow the growth rates of two critical modes to be controlled
independently.
The linear part of the PDE~$\cal{L}$ acts on a mode $e^{ikx}$ with
eigenvalue~$\sigma(k)$.
The dependence of the eigenvalue on~$k$ is specified by
$\sigma(1)=\mu$ and $\sigma(q)=\nu$, controlling the growth rates of the modes
of interest; $\sigma'(1)=\sigma'(q)=0$; 
and $\sigma(0)=\sigma_0<0$, controlling the depth of the minimum between
$k=1$ and $k=q$. 
With $\sigma$ an even function of~$k$,
these requirements lead to a fourth-order polynomial in~$k^2$:
 \begin{equation*} 
 \sigma(k)=\frac{k^2\left(A(k)\mu+B(k)\nu\right)}{q^4(1-q^2)^3}
           + \frac{\sigma_0}{q^4}(1-k^2)^2(q^2-k^2)^2,
 \end{equation*} 
where $A(k)=(k^2(q^2-3)-2q^2+4)(q^2-k^2)^2q^4$ and
      $B(k)=(k^2(3q^2-1)+2q^2-4q^4)(1-k^2)^2$. 
The linear operator~${\cal{L}}$ is obtained by replacing $k^2$ by~$-\nabla^2$.
The nonlinear terms in the PDE model are simple quadratic and cubic
combinations of $U$ and its derivatives. Standard weakly nonlinear theory
gives the values of $\Qzz$, $\Qzw$, $\Qww$ and $\Qwz$: having the three
quadratic terms in (\ref{eq:PDE}) enables different sign combinations to be
chosen in the amplitude equations (\ref{eq:zzw}) and~(\ref{eq:zww}).

% $\Qzz = 2Q_1 - 2 Q_2 + (2-q^2) Q_3$,
% $\Qzw = 2Q_1 - (1+q^2) Q_2 + q^2 Q_3$,
% $\Qww = 2Q_1 - 2 q^2 Q_2 + (2 q^2-1) Q_3$ and 
% $\Qwz = 2Q_1 - (1+q^2) Q_2 + Q_3$, so 

%%%%%%%%%%%%%%%%%%%%%%%%%%%%%%%%%%%%%%%%%%%%%%%%%%%%%%%%%%%%%%%%%%%%%%%%%%%%%%%%
% Unfolding diagrams
\begin{figure}
\def\framebox#1{{#1}}%
\makebox[\hsize]{%                                                              
\hfil\framebox{\raisebox{2.20truein}{(a)}}%
\framebox{\includegraphics[width=0.74\hsize]{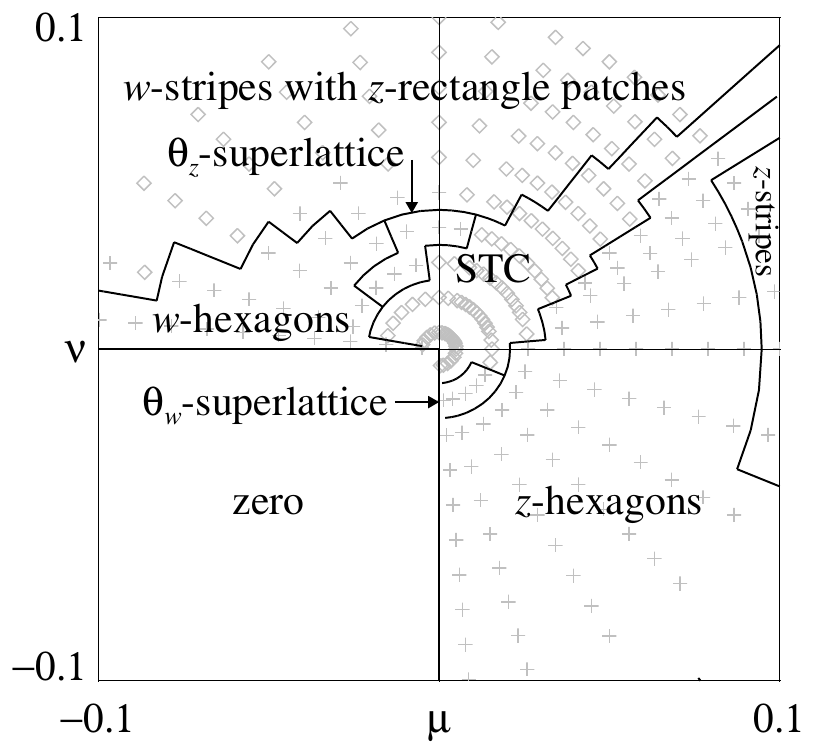}}%
\hfil}

\vspace{0.2ex}

\makebox[\hsize]{%                                                              
\hfil\hfil\framebox{\raisebox{1.22truein}{(b)}}\hfil%
\framebox{\includegraphics[width=0.39\hsize]{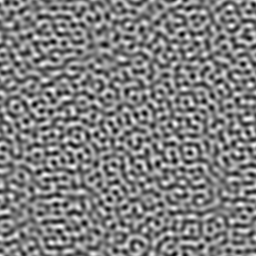}}%
\hfil\hfil\framebox{\raisebox{1.22truein}{(c)}}\hfil%
\framebox{\includegraphics[width=0.39\hsize]{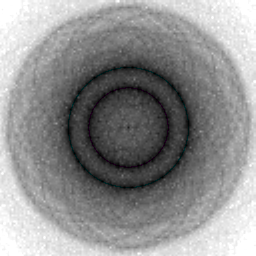}}%
\hfil\hfil}

\vspace{-2.0ex}

\caption{\label{fig:unfolding}(a) Bifurcation set showing the patterns that are
seen for $q=0.66$, $\sigma_0=-2$, $Q_1=0.3$, $Q_2=1.3$, $Q_3=1.7$ (so
$\Qzz\Qzw<0$ and $\Qww\Qwz<0$). 
We find $z$-hexagons ($k=1$),
$w$-hexagons ($k=q$), spatio-temporal chaos
(STC), 
mixed patterns ($w$-stripes with patches of $z$-rectangles) and 
two types of superlattice pattern: one with 6 modes on the outer circle
and 12 on the inner,
separated by~$\theta_w$, and the other with 6 modes on the inner circle
and 12 on the outer,
separated by~$\theta_z$. 
Steady (time-dependent) patterns are indicated with a~$+$ ($\diamond$). 
The $(\mu,\nu)$ scale is not
uniform.
 (b)~Pattern for parameters in the STC region,
with $\mu=\nu=0.00707$. The correlation length of this pattern is about
$1$--$2$~wavelengths.
 (c)~The two critical circles are clearly seen
in its power spectrum.}
 \end{figure}
%%%%%%%%%%%%%%%%%%%%%%%%%%%%%%%%%%%%%%%%%%%%%%%%%%%%%%%%%%%%%%%%%%%%%%%%%%%%%%

Simulations of equation (\ref{eq:PDE}), carried out on a $30\times30$ domain for a range of $(\mu,\nu)$, are
shown in figure~\ref{fig:unfolding}(a) for $q=0.66$ and values of $Q_1$, $Q_2$ and $Q_3$ that give
$\Qzz\Qzw<0$ and $\Qww\Qwz<0$. We find a variety of patterns and STC, with a typical solution shown in
figure~\ref{fig:unfolding}~(b,c). For other choices of $q$ and the quadratic coefficients, the behavior of
the PDE is broadly in line with the expectations. When $q<\frac{1}{2}$, we find no examples of~\hbox{STC}.
For $q>\frac{1}{2}$, we find steady and time-dependent patterns with many modes on both critical circles. In
general, with both $\Qzz\Qzw$ and $\Qww\Qwz$ positive, these patterns are all steady; time dependence (chaos
or STC) is much more common when both $\Qzz\Qzw$ and $\Qww\Qwz$ are negative. We find steady and
time-dependent quasipatterns only with the special values $q=0.5176$ and $q=0.6180$. Further details will be
presented elsewhere.

We conclude that whenever there are 3WIs between waves on two critical circles with radius ratio between
$\frac{1}{2}$ and $2$, interactions in both directions must be taken into account. This is generic: these
two-way 3WIs will be important in any problem with two comparable length scales. Most values
of~$q>\frac{1}{2}$ lead to the possibility of generating an infinite number of interacting waves. The
exceptions are those values associated with 10- and 12-fold quasipatterns. The outcome of the 3WIs will be
influenced by the signs of the quadratic coefficients in (\ref{eq:zzw}) and (\ref{eq:zww}), with
time-dependence and STC most likely in the case of both pairs of quadratic coefficients having opposite sign.
These ideas have been confirmed in numerical investigations of a model PDE~(\ref{eq:PDE}). By computing
quadratic coefficients from the Navier--Stokes equations, we have shown that these ideas are in broad
agreement with the experimental results of~\cite{Ding2006}: as the phases in the forcing are varied, switches
between 12-fold quasipatterns or $22^\circ$ superlattice patterns and STC line up with changes of sign of the
products of quadratic coefficients.
 % \looseness=-1

\begin{acknowledgments}
We acknowledge valuable conversations with Yu Ding, Jay Fineberg, Paul
Umbanhowar and Jorge Vi\~{n}als,
as well as comments from the referees.
MS is grateful for support from the National
Science Foundation (DMS-0709232).
 \end{acknowledgments}

% \bibliography{rss}% Produces the bibliography via BibTeX.   

\end{document}